\documentclass[12pt]{article}
\usepackage{amsmath}
\usepackage{psfig,epsfig}
\usepackage{latexsym}
\usepackage{pifont}

\newcommand{\lapprox}{\lower.6ex\hbox{$\; \buildrel<\over\sim \;$}}

\newcommand{\gapprox}{\lower.6ex\hbox{$\; \buildrel>\over\sim \;$}}

\newcommand{\curly}{\lower1.ex\hbox{$\; \stackrel{\textstyle \wr}{\wr} \;$}}

\def\barr{\begin{array}}
\def\earr{\end{array}}
\def\berr{\begin{eqnarray}}
\def\err{\end{eqnarray}}
\def\berrno{\begin{eqnarray*}}
\def\errno{\end{eqnarray*}}
\def\be{\begin{equation}}
\def\ee{\end{equation}}

\def\fr{\frac}

\def\apj{{\it Astrophys.~J.}}

\def\prd{{\it Phys.~Rev.~D.}}

\def\mnras{{\it Mon. Not. R.~Astr.~Soc.}}

\def\pl{{\it Phy. Lett.}}

\renewcommand{\a}{\alpha}

\renewcommand{\L}{\Lambda}

\renewcommand{\t}{\theta}

\title{\bf Cosmological Constraints on a Power Law Universe}
\author{Geetanjali Sethi\thanks {E-mail: getsethi@physics.du.ac.in}\\
       { Department of Physics and Astrophysics,} \\
       { University of Delhi, Delhi-110007, India}\\
\\
{Abha Dev} \\
{Miranda House, University of Delhi, Delhi 110 007}\\
and\\
Deepak Jain \\
       { Deen Dayal Upadhyaya College,University of Delhi, Delhi 110 015 } \\
       }
\begin{document}

\maketitle
 
\begin{abstract}

Linearly coasting cosmology is comfortably concordant with a host of
cosmological observations.  
It is surprisingly an  excellent fit to  SNe Ia observations and
constraints arising from age of old quasars.
In this article we highlight the overall viability of an open linear coasting cosmological model. 
The model is consistent with the latest SNe Ia ``gold''
sample and accommodates a very old high-redshift quasar, which the
standard cold-dark model fails to do. 

\end{abstract} 

\section{Introduction}

In the past there has been a spurt of activity to explain the observed
``accelerated expansion'' of the universe. Classes of $\Lambda$CDM
models as well as quintescence models have been designed to
accommodate such an expansion deduced from 
observations on high-redshift Supernovae Ia (SNe Ia) \cite{perl}.
 
The SNe Ia look fainter than they are expected in the
standard Einstein-De Sitter model, which was the favoured model
prior to these observations. 
In standard cosmology, these results, when combined with
the latest CMB data and clustering estimates, are used to make out a
case for a universe in which  accelerated expansion is
fueled by a self-interacting, unclustered fluid, with high negative
pressure, collectively known as $''Dark~ Energy''$ (for latest review
see \cite{varun}), the simplest and the most favoured candidate being
the cosmological constant ($\L$). Consequently, several models with a relic
cosmological constant $(\L CDM)$, have been used to best describe the 
observed universe. However, most of them suffer from severe
fine tuning problems \cite{varun,weinberg}. The basic reason is the
wide spread belief that the early universe evolved through a cascade
of phase transitions, thereby yielding a vacuum energy density which
is presently 120 orders of magnitude 
smaller than its value at the Planck time. Such
a discrepancy between theoretical expectations and empirical
observations constitute a fundamental problem  at interface of
astrophysics, cosmology and particle physics. In the last few years,
several attempts have been made to alleviate the cosmological
constant problem. For example, in the so called dynamical $\L(t)$
scenarios (or deflationary cosmology), the cosmological term is a
function of time and its presently observed value is a remnant of
primordial inflationary/deflationary stage \cite{ozer}. Other examples
are scenarios in which the evolution of classical fields are coupled
to the curvature of the space-time background in such a way that their
contribution to the energy density self-adjusts to cancel the vacuum
energy \cite{models}, as well as some recent ideas of a SU(2)
cosmological instanton dominated universe \cite{allen}. At least in
the two later examples, an interesting feature is a power-law growth
for the cosmological scale factor $a(t) \approx t^{\a}$, where $\a$ 
may be constrained by observations.

In a series of eariler articles, we have explored the viability of a model
that has $a(t) \approx t^{\a}$ with $\a \gapprox 1$
\cite{meetu,abha,annu,savita,geetanjali}. 
The motivation for such an endeavor comes from several
considerations. Such models do not
have a horizon problem. Moreover, the scale factor in such theories
does not constrain the density parameter and therefore, they are free
from flatness problem. There are also observational motivations for
considering power-law cosmologies. For $\a \ge 1$, the predicted
age of the universe is $t_{0} \ge H_{0}^{-1}$, i.e. at least
$50 \%$ greater than the prediction of the standard flat
model (without cosmological constant). This makes the universe
comfortably in agreement with the recent age estimates of globular
clusters and high-z redshift galaxies. 

A linear evolution of the scale factor is supported in some
 alternative gravity theories \cite{meetu}, as well as in standard
 model with a specially chosen equation of state \cite{kolb}. It was
 reported by Abha Dev et al. \cite{abha} that this model is consistent
 with gravitational lensing statistics (within $1\sigma$ level) and
 the constraints from the ages of old high redshift galaxies. It was
 also demonstrated that this model is consistent with primordial
 nucleosynthesis \cite{annu}. For $\Omega = 0.65$ and $\eta = 7.8
 \times 10^{-9}$, the model with $\a = 1$ yields ${He}^{4} = 0.23$ and
 metallicity of the range $10^{-7}$ \cite{geetanjali}. 
 Linear coasting surprising clears preliminary constraints on structure
 formation and CMB 
 anisotropy \cite{savita}.      

In this article we explore the concordance  of an open linear coasting
model with the latest SNe Ia data and bounds from age estimates 
of old quasars.In Section 2, we give
the basic equations for the model adopted. In Section 3.1, we
constrain parameter $\a$ by using SNe Ia ``Gold Sample''. The lower
 bound on $\a$ from the age
estimates of an old high redshift quasar is discussed in Section
 3.2. We summarize the results in Section 4.

\section{Linear Coasting Cosmology}

We consider a general power law cosmology with the scale factor given in
terms of an arbitrary dimensionless parameter $\a$
\be
a(t)= \frac{c}{H_{0}}\left(\fr{t}{t_0}\right)^{\a}\,\,
\label{eq:ansatz}
\end{equation}
for an open FRW metric  
\be
ds^2=c^2dt^2-a^2(t) \left[\frac{dr^2}{1 + r^2} + r^2(d\t^2
+\sin^2{\t}\,d\phi^2)
\right]\,\,.
\end{equation}
Here $t$ is cosmic proper time and $r,\,\t,\,\phi$ are comoving spherical 
coordinates. 

The expansion rate of the universe is described by a Hubble parameter, $H(t) =
\dot{a}/a ={\a}/{t}$. The {\it present} expansion rate of the universe is
defined by a Hubble {\it constant}, equal in our model to $H_0=\a/t_0$ (here
and subsequently the subscript 0 on a parameter refers to its present value).
The scale factor and the redshift are related to their present values by
$a/a_0=(t/t_0)^{\a}$.  As usual, the ratio of the scale factor at the emission
and absorption of a null ray determines the cosmological redshift $z$ by
\be
\frac{a_0}{a(z)} = 1+z\,\,,
\label{eq:redshift}
\end{equation}
and the age of the universe is
\be
t_z=\fr{\a}{H_0(1+z)^{1/\a}}\,\,.
\label{eq:age}
\end{equation}
Using (\ref{eq:redshift}), we define the dimensionless Hubble parameter   
\be
h(z) \equiv \frac{H(z)}{H_0}=(1+z)^{1/\a}.
\label{eq:dimensionless_hubble}
\end{equation}
The present `radius' of the universe is defined as (see Eq.~\ref{eq:ansatz}) 
\be
a_0 =\frac{c}{H_{0}}\,\,.
\label{eq:a_0}
\end{equation}

\noindent In terms of the parameter $\a$, the luminosity distance between two redshifts $z_{1}$ and $z_{2}$ is 
\be
d_{\rm L}(z_{1},z_{2}) = \fr{c(1+z_{2})}{H_{0}} \times
\sinh\left[\fr{\a}{\a-1}\left
  \{(1+z_{2})^{\fr{\a-1}{\a}}-(1+z_{1})^{\fr{\a-1}{\a}}\right \}
  \right]\,\,. 
\label{DL}
\end{equation} 

\noindent In a limiting case,  $\a \rightarrow 1$, we obtain 
\be
d_{\rm L}(z_{1},z_{2}) =
\fr{c}{2H_{0}}\fr{[(1+z_{2})^{2}-(1+z_{1})^{2}]}{(1+z_{1})}\,\,.
\end{equation}

\noindent The look-back time, the difference between the age of the universe
when a particular light ray was emitted and the age of the universe
now, is
  
\be
\fr{c\,dt}{dz_{\rm L}}=\fr{c}{H_0 (1+z_{\rm L})^{\fr{\a
+1}{\a}}}\,\,. 
\label{eq:backtime}
\end{equation}    

\section{The Observational Tests}
\subsection{ Constraints from SNe Ia data}

Of late, properties of type Ia supernovae (SNe Ia) as  excellent 
cosmological standard candles has elevated the status of the Hubble 
flow to that of a precision measurement.
The magnitude of a ``standard candle'' is related to its
luminosity distance $d_L$ through 
\be
m(z) = M + 5\,\log_{10}\,\left [{d_L\over {\mathrm{Mpc}}}\right ] + 25,
\ee

\noindent where $M$ is the absolute magnitude and is assumed to be
constant for a standard candle like SNe Ia. The apparent magnitude can
also be expressed in terms of dimensionless luminosity distance
$\mathcal D_L(z)$ as:

\be
m(z) = {\mathcal M} + 5 \,\log_{10}{\mathcal D_L(z)}\,\,,
\ee

\noindent with
\be
{\mathcal D_L(z)}=\fr{H_{0}}{c} d_{\rm L}
\ee
and 
{\setlength\arraycolsep{2pt}
\berr
{\mathcal M} & = & M +5 \,\log_{10}\left( \fr {c/H_{0}}{1 Mpc}\right) + 25  \nonumber \\
& = & M - 5 \log_{10}h + 42.38 \,\,.
\err}

For our analysis we use ``gold'' sample compiled by Reiss et al.\cite{reiss}.
The sample consists  of 157 data points which are in the terms of
distance modulus

\begin{eqnarray}
\mu_{obs}&= &m(z) -{ M}\nonumber \\
 & = &5 \,\log_{10}{\mathcal D_L(z)}- 5 \log_{10}h + 42.38
 \end{eqnarray}

\noindent The best fit model to the observations is obtained by using 
$\chi^2$ statistics i.e. 
\be
\chi^{2} =
\sum_{i=1}^{157} \left [ \frac{\mu_{th}^{i}-\mu_{obs}^ i}
{\sigma_i} \right ]^{2},
\ee

\noindent where $\mu_{th}$ is the predicted distance modulus for a
supernova at redshift z and $\sigma_i$ is the dispersion of the
measured distance modulus due to intrinsic and observational
uncertainties in SNe Ia peak luminosity. In order to integrate  over
the Hubble constant, we  use  the modified $\chi^2$
statistics as defined in the ref.\cite{wang} 

\be
{\bar\chi^{2}} = \chi_{*}^{2} - {C_1\over C_2}
\left (C_1 + {2\over 5}\ln 10\right ) - 2 \ln h^{*}.
\ee

Here $h^{*}$ is the fiducial value of the dimensionless Hubble constant. 
And
\be 
\chi_{*}^{2}\equiv
\sum_{i} \left [ \frac{\mu_{th}^{{*}(i)}-\mu_{obs}^ i}
{\sigma_i} \right ]^{2},
\ee

\be 
C_1 \equiv \sum_{i}  {[{\mu_{th}^{{*}(i)}-\mu_{obs}^ i}]\over
{{\sigma_i}^{2}}},
\ee

\be 
C_2 \equiv \sum_{i}  {1\over
{{\sigma_i}^{2}}},
\ee

\noindent with 
\be
\mu_{th}^{{*}(i)}(z_i, h = h^{*})= 5 \,\log_{10}{\mathcal D_L(z)}- 5
\log_{10}h + 42.38. 
\ee

\noindent For our calculations, we use $h^* = 0.72$. 
We work with the following range of the parameter $\a$: $0.0 \leq \a
\leq 3.0$. We perform a grid
search in the parametric space to find the best fit model.
For a one parameter fit, the 68\% Confidence Level (CL) (90\% CL)
corresponds to $\Delta\bar\chi^2 = 1.0$ ($2.71$). 

Figure 1 shows variation of $\bar\chi^2$ with $\a$. We find that the
 minimum of $\bar\chi^2$  i.e $\bar\chi^2_{min}$occurs for $\a =
 1.04$, with $\bar\chi^2_{\nu} = 
 1.23$ ($\bar\chi^2_{\nu} = \bar\chi^2_{min}$ / degree of freedom).
The SNe Ia  data thus provides the following constraints: 
$0.98 \leq \a \leq 1.11$ at 68\% CL and $0.95 \leq \a \leq 1.15$
at 90\% CL. 

\subsection{Constraints from Age Estimates of an Old, High-z Quasar}

The age estimates of old high redshift objects play a very important 
role in constraining cosmological parameters \cite{al}.
The recently discovered  quasar APM 08279+5255 at a
redshift of  z = 3.91 is very important object in this regard
\cite{has}. Conservative estimates of its age have been made
from iron enrichment in detailed chemodynamical modelling and give 
a staggering value of at least 2 Gyr for this object.
Standard flat FRW models with cosmological constant  fail to accommodate
this old, high redshift quasar\cite{fri}. In this letter we use
this quasar  to put limits on the $\a$ in power law cosmologies. 
\vskip 0.5cm

\noindent The age-redshift relationship in power-law cosmology is
given as $$t_z = {\alpha\over {H_0( 1+z)^{1/\alpha}}}.$$ 
In order to constrain  $\a$ from the age estimate of the above
mentioned quasar we follow ref.\cite{al}. The age of the universe at a 
given redshift has to be  greater than or at least equal to the age 
of its oldest objects at that redshift. In a power law cosmology 
the age of the universe increases 
with increasing $\a$. Hence this test provides lower bound on $\a$. This 
can be checked if we define the dimensionless ratio:
\be
{t_z\over t_{\rm q}} = {f(\alpha, z)\over T_q = H_0 t_{\rm q}} \ge 1\,\,.
\ee
where $t_{\rm q}$ is the age of an old object (here the quasar) at a
given redshift and
$f(\a, z)=\a/(1+z)^{1/\a}$, is a dimensionless factor. For every
high redshift object, $T_q = H_0t_q$ is a dimensionless age parameter. 
The error bar
on $H_{0}$ determines the extreme value of $T_{\rm q}$. The lower  
limit on $H_0$ is updated to nearly $10\%$ of accuracy by Freedman
\cite{fre}: $H_0 = 72 \pm 8$ km/sec/Mpc. So the 2 GYr old quasar
at z = 3.91 gives $T_q = 2.0 H_0$ Gyr and hence $0.131 \le T_q \le
0.163$. We use minimal value of the Hubble constant, $ H_0= 64$ km/sec/Mpc, to
get strong conservative limit. It thus follows that  $T_q \ge
0.131$. Only those values of $\a$ are allowed for which the age of
the universe at z = 3.91 equals to or is  greater than the age of the
quasar at that redshift i.e $H_0 t_z(z=3.91)\ge H_0t_q$.

Figure 2 shows the variation of dimensionless age parameter 
$H_0t_z(z=3.91)$ as a
function of $\a$. The horizontal line in the Figure corresponds to the
age of the quasar which is $ T_q = 0.131$. It is clear from
the Figure that $\a$ should be at least 0.85 in order to allow this
quasar to exist in  power law cosmology.

\section{Discussions}

\begin{table}
\begin{center}
\begin{tabular}{l l r}\hline\hline
Method &  Reference & $\a$  \\          
\hline
\hline
Lensing Statistics &  & \\
(Optical Sample)& & \\
(i) $n_{\rm L}$ & Dev et al.\cite{abha}&  $1.09 \pm 0.3$ \\
(ii) Likelihood &Dev et al.\cite{abha}  & $1.13^{+0.4}_{-0.3}$\\
\hspace{0.5cm} Analysis  &  &   \\
& & \\
OHRG & Dev et al.\cite{abha} & $\ge 0.8$ \\
& & \\
SNe Ia & This  Letter & $1.04^{+0.07}_{-0.06}$ \\
(Gold Sample)& & \\
&&\\
Old Quasar & This Letter & $\ge 0.85$\\
\hline 
\end{tabular}
\caption{Constraints on $\a$ from various cosmological tests.}
\label{T2}
\end{center}
\end{table}

Recent observations of Type Ia supernovae lead to the discovery of
an accelerating universe. This acclerated expansion has been
attributed to a dark energy component with high negative pressure. The
simplest model for dark energy is the cosmological constant that comes
with its theoretical and fine tuning problems.
We have been exploring alternative  models of universe
which have the potential of explaining these observations. 

The main results of this article along with the constraints obtained
from the gravitational lensing statistics of the optical sample and 
age estimates of old high redshift galaxies \cite{abha} are summarized
in Table 1. The motivation 
for our work was to establish the viability of a linear coasting
cosmology $a(t) = t$. Using Sne Ia data, we find that such a model is
well accommodated within 1 $\sigma \,$: $0.98 \leq \a \leq 1.15$.
The age estimates of the old, high redshift quasar at $z = 3.91$ 
give the lower bound $\a \ge 0.85$ for a power law cosmology.          
We thus find that $\a = 1.0$ is in concordance with the 
observational tests listed in Table 1. It is interesting 
to observe that the SNe Ia data and the age estimates of the old, 
high redshift quasar rule out an Einstein-de Sitter universe ($ \a = 2/3$).
We conclude that the coasting cosmology with 
strictly linear evolution of scale factor, $a(t) = t$, is in excellent 
agreement with these observations.

\section*{Acknowledgments}

The authors are grateful to Daksh Lohiya
for useful discussions during the course of this work.

\pagebreak
 
\begin{figure}[ht]
\vspace{-1.2in}
\centerline{
\epsfig{figure=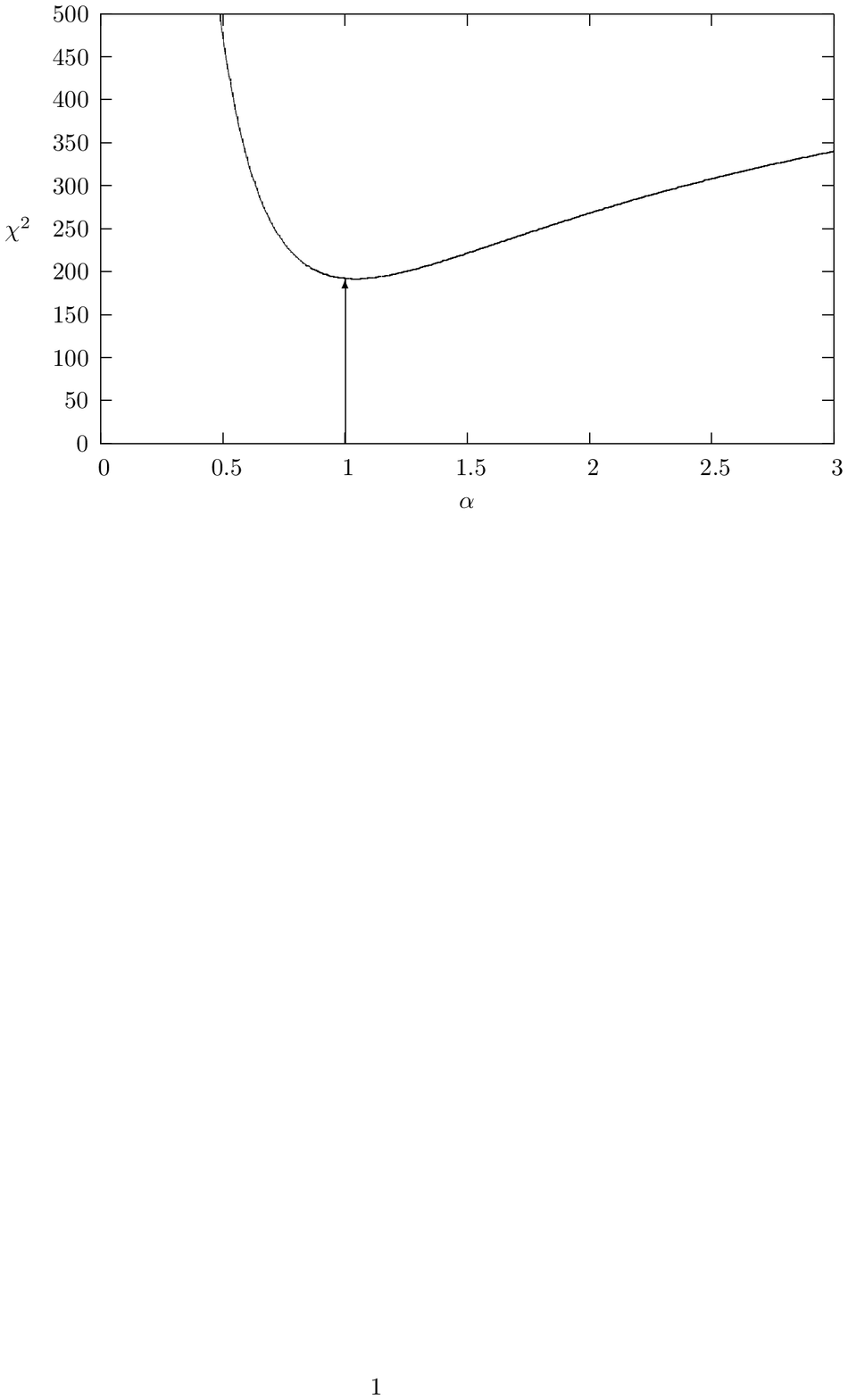,width=1.5\textwidth}}
\vspace{-7.2in} 
\caption{Variation of ${\bar\chi^{2}}$ with $\a$. The arrow corresponds
  to the minimum value of ${\bar\chi^{2}}$ which occurs for $ \a =
  1.04$\;.}

\end{figure}

\begin{figure}[ht]
\centerline{
\epsfig{figure=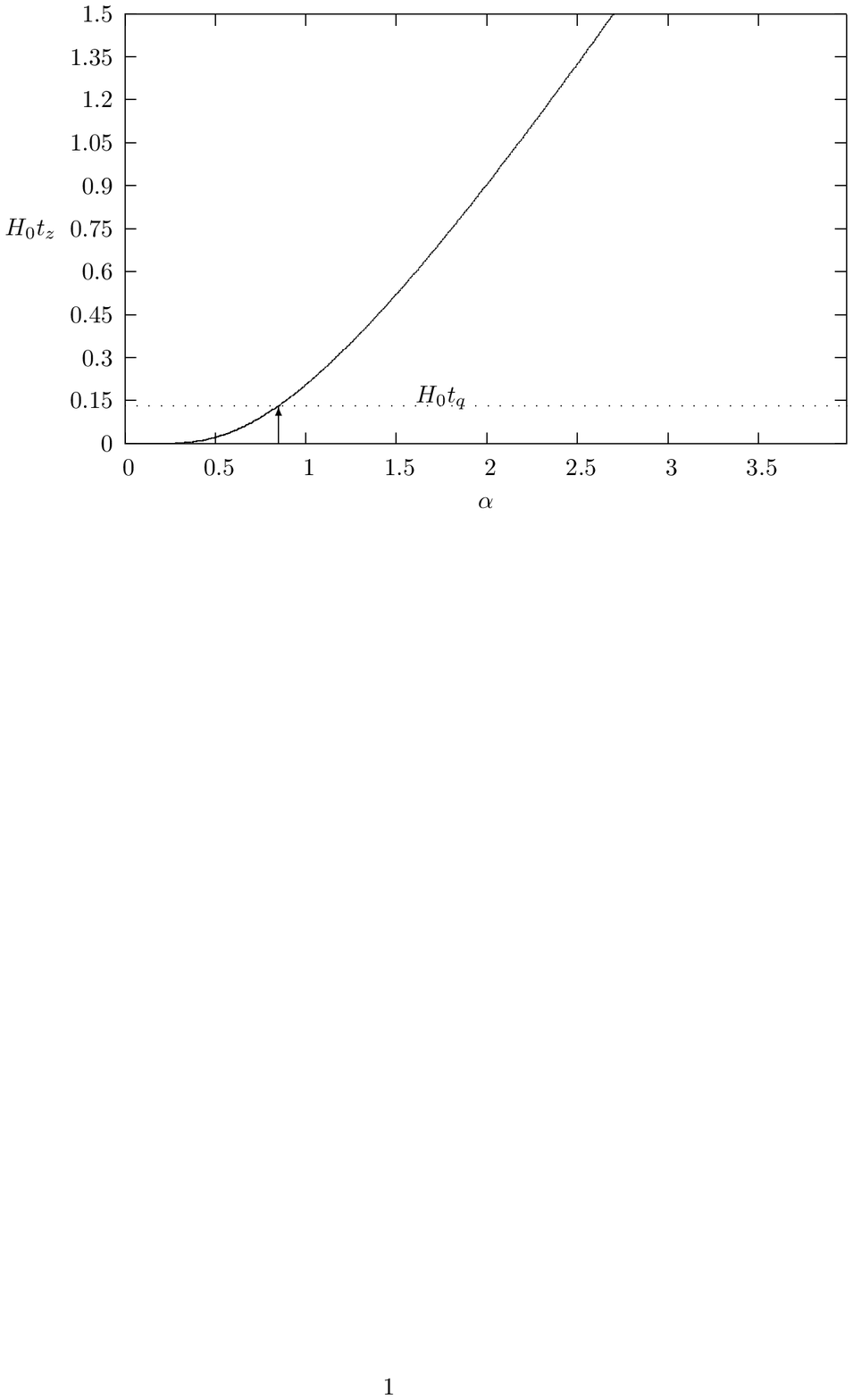,width=1.5\textwidth}}
\vspace{-7.2in} 
\caption{Variation of dimensionless age parameter, $H_0t_z$, as a
  function of $\a$ for $z = 3.91$. The dotted line corresponds to the
  dimensionless age parameter, $H_0t_q$, of the old quasar at this
  redshift. It is clear that the lower bound on value of $\a$ to
  accomodate the quasar is 0.85.}
\end{figure}

\end{document}